\documentclass[twocolumn]{aastex631}

\usepackage{newtxtext,newtxmath}

\newcommand{\mathbfit}[1]{\textbf{\textit{#1}}}

\begin{document}

\title{YSO Jets Magnetocentrifugally Driven by Reconnecting Atmospheric Avalanche Accretion Streams Above Inner Circumstellar Disks}

\author[0000-0003-2929-1502]{Yisheng Tu}
\affiliation{Astronomy Department, University of Virginia, Charlottesville, VA 22904, USA}

\author{Zhi-Yun Li}
\affiliation{Astronomy Department, University of Virginia, Charlottesville, VA 22904, USA}

\author{Zhaohuan Zhu}
\affiliation{Department of Physics and Astronomy, University of Nevada, Las Vegas, NV, 89154-4002, USA}

\author{Xiao Hu}
\affiliation{Astronomy Department, University of Virginia, Charlottesville, VA 22904, USA}
\affiliation{Department of Astronomy, University of Florida, Gainesville, FL 32608, USA}

\author{Chun-Yen Hsu}
\affiliation{Astronomy Department, University of Virginia, Charlottesville, VA 22904, USA}



\begin{abstract}

Fast, collimated jets are ubiquitous features of young stellar objects (YSOs). They are generally thought to be powered by disk accretion, but the details are debated. Through 
2D (axisymmetric) MHD simulations, we find that a fast ($>100$~km/s) collimated bipolar jet is continuously driven along the north and south poles of the circumstellar disk that is initially magnetized by a large-scale open poloidal field and contains a thermally ionized inner magnetically active zone surrounded by a dead zone. The fast jet is primarily driven magneto-centrifugally by the release of the gravitational binding energy of the so-called ``avalanche accretion streams" near the boundary of an evacuated poloidal field-dominated polar region and a thick disk atmosphere raised by a toroidal magnetic field. Specifically, the fast outflow is driven along the upper (open) branch of the highly pinched poloidal field lines threading the (strongly magnetically braked) accretion streams where the density is relatively low so that the lightly loaded material can be accelerated magneto-centrifugally along the open field line to a high speed. The highly pinched poloidal magnetic fields threading the avalanche accretion streams tend to reconnect, enabling mass to accrete to the center without dragging along the poloidal magnetic flux with it. The reconnection provides a potential heating source for producing chondrules and calcium- and aluminum-rich inclusions (CAIs).
\end{abstract}



\section{Introduction}

Jets are observed in many astrophysical systems, from young stellar objects (YSOs) to Active Galactic Nuclei (AGNs). The mechanisms that drive these observed jets and whether all jets share the same mechanism are areas of active debate. AGNs can rely on the spinning black hole itself to power the jet electromagnetically \citep[][]{Blandford1977}. YSOs do not have such a luxury and must instead rely on the release of the gravitational binding energy of the circumstellar disk accretion for jet launching, typically through magnetic fields threading the disk.   

Depending on where the jet-launching magnetic fields are rooted on the disk, there are two broad classes of models: the X-wind \citep[][]{Shu2000, Shang2007} and disk-wind \citep[][]{Konigl2000, Pudritz2007} models. The former envisions the wind-launching open field lines concentrating near the corotation radius where the stellar magnetosphere truncates the circumstellar disk \citep[see, e.g.][]{Ostriker1995, Zanni2009, Zanni2013},
potentially carrying away the angular momentum associated with the mass accretion along (deformed) magnetospheric field lines (the so-called ``funnel flow") onto the central star, preventing excessive stellar spin-up. 
The latter class of models posits that the wind-launching field lines are anchored over a range of disk radii. It has the potential to remove angular momentum and drive disk accretion over a wide range of radii. 




Our work focuses on the magnetically driven disk-wind model in the YSO context, which has a long and rich history, as reviewed by, e.g., \citet[][and reference therein]{Pudritz2007} and more recently \citet{Ray2021}. A key issue of the magnetic disk-wind theory is the origin and distribution of the wind-launching large-scale poloidal magnetic field on the disk. Since the YSO disks are formed out of dense cores of molecular clouds that are now known to be magnetized by ordered fields \citep[][]{Pattle2023}, it is reasonable to expect some of the core magnetic flux to be dragged into the disk during its formation stage by the protostellar envelope collapse \citep[e.g.][]{Tu2024a, Mauxion2024}; there is, therefore, no shortage of poloidal magnetic flux that can in principle thread the disk and drive the disk wind. 

What fraction of the core's magnetic flux is inherited by the disk and the subsequent flux evolution on the disk are important \citep[e.g.][]{Yang2021} but open questions. In particular, the wind-driven radial mass accretion has the tendency to drag the wind-launching poloidal magnetic flux inward and concentrate it near the inner edge of the disk, potentially disrupting the disk, which may, in turn, shut off the wind launching. This magnetic flux problem for jet/disk-wind launching in a disk threaded by a net poloidal flux is analogous to the classic ``magnetic flux problem" in star formation \citep[][]{Mestel1956}. 

How the magnetic flux problem in jet/disk-wind launching is resolved remains uncertain. In the lightly ionized outer disk, ambipolar diffusion can allow disk material to accrete across the wind-launching poloidal field lines without dragging them along, as demonstrated semi-analytically \citep[e.g.][]{Wardle1993, Li1996} and through numerical simulations \citep[e.g.][]{Suriano2018, Martel2022}. The situation is less clear in the inner disk, where the temperature is high enough that the gas is well coupled to the magnetic field through the thermal ionization of alkali metals and where the fast jet is expected to be launched. 

Previous semi-analytic work and numerical simulations of magnetic disk winds from the magnetically well-coupled inner disks have typically assumed a prescribed resistivity, presumably of turbulent origin \citep[e.g.][]{Ferreira1993, Li1995, Ferreira1997, Casse2002, Tzeferacos2009, Murphy2010, Stepanovs2016}, which facilitates radial disk accretion across poloidal field lines, thus alleviating the magnetic flux problem. In particular, \cite{Stepanovs2016} demonstrated that resistive disks can drive fast steady magneto-centrifugal outflows over a wide range of disk magnetization. Whether such disks can drive fast jets persistently without a prescribed resistivity remains less explored. It is the main goal of our paper. 

There have been several recent simulations of magnetically well-coupled inner disks without an explicitly prescribed resistivity, with a stellar magnetosphere \citep[e.g.,][]{Lubow1994, Pantolmos2020, Ireland2021, Takasao2022, Zhu2024, Meskini2024} or without \citep[e.g.][]{Tzeferacos2009, Fendt2013, Zhu2018, Takasao2018, Mattia2020a, Mattia2020b, Jacquemin-Ide2021}. 
These simulations show that the fast accretion of the thick magnetically dominated surface can counterbalance the field's outward diffusion. As a first step, we will ignore the stellar magnetosphere and focus on the question of whether a fast jet can be driven by the inner disk alone and, if so, what mechanism is responsible for the jet launching and the question of whether the magnetic flux problem can be resolved in the absence of any explicit resistivity. We find that a persistent fast jet is driven magneto-centrifugally along the lightly mass-loaded upper (open) branch of the highly pinched poloidal field lines threading the so-called ``avalanche accretion streams" in the elevated disk atmosphere and that the magnetic flux problem is alleviated by the reconnection of the highly pinched field of jet-launching field lines threading the accretion streams. 

The paper is organized as follows. The simulation setup and results are described in \S~\ref{sec:methods} and \S~\ref{sec:result}, respectively. We discuss the results in \S~\ref{sec:discussion} and conclude in \S~\ref{sec:conclusion}.

\section{Simulation Setup}
\label{sec:methods}

The launch of a jet from the inner circumstellar disk is a complex process involving non-ideal magneto-hydrodynamics. In particular, Ohmic dissipation is important in the relatively cool dead zone of the disk surrounding the hotter inner active zone, where the gas is close to the ideal MHD limit due to thermal ionization. The system is governed by the following set of equations: 
\begin{equation}
    \frac{\partial\rho}{\partial t} + \nabla\cdot(\rho\mathbfit{v}) = 0,
\end{equation}
\begin{equation}
    \rho\frac{\partial \mathbfit{v}}{\partial t} + \rho(\mathbfit{v}\cdot\nabla)\mathbfit{v} = -\nabla P + \frac{1}{c}\mathbfit{J}\times\mathbfit{B} + \rho \mathbfit{g},
    \label{equ:mhd momentum equation}
\end{equation}
\begin{equation}
    \frac{\partial \mathbfit{B}}{\partial t} = \nabla\times(\mathbfit{v}\times\mathbfit{B}) - \frac{4\pi}{c}\nabla\times[\eta_O (\mathbfit{J}\times\hat{\mathbfit{B}})],
\end{equation}
where $\mathbfit{J} = (c/4\pi)\nabla\times\mathbfit{B}$ is the current density, $\eta_O$ the Ohmic diffusivity, and $\mathbfit{g}$ is the gravitational acceleration from the central star, which we assume to have one solar mass. Other symbols have their usual meanings. 

The governing equations are solved using the ATHENA++ code \citep{Stone2020} with static mesh refinement (SMR) in spherical polar coordinates. The simulation domain extends from 0.03 au to 10 au in the radial direction and $0.02$ rad to $\pi - 0.02$ rad in the polar direction. The domain is chosen to cover both the magnetically active and dead zones on the disk and the atmosphere and outflow outside the disk. The base resolution has 160 logarithmically spaced cells in the radial direction and 192 uniformly spaced cells in the polar angle direction. To increase the resolution of the jet-launching inner region, one level of static mesh refinement (SMR) is added from the inner boundary to $5$ au in the radial direction and from the north polar axis to the south polar axis in the polar direction within the 5 au radius.

For illustration purposes, we adopt a power-law distribution for the initial midplane density: 
\begin{equation}
   \rho_\mathrm{mid}(r) = \rho_0\Big(\frac{r}{r_0}\Big)^{p},
\end{equation}
 The vertical density distribution within a dimensionless height of $\vert z/r\vert =0.1$ (where $z$ is the vertical height from the disk midplane) has a profile set by hydrostatic equilibrium. We continue the same profile to larger heights (where $\vert z/r\vert > 0.1$). To help with numerical stability, particularly at small radii and in the polar regions where the magnetic field tends to be strong, we adopt a radially varying density floor:
\begin{equation}
    \rho_\mathrm{floor}(r) = 
    \begin{cases}
      \rho_f ;             & r \leq r_f, \\
      \rho_f (r / r_f)^{-2}; & r > r_f,,
    \end{cases}
\end{equation}
where $\rho_f = 5.94\times10^{-18}$~g cm$^{-3}$ and $r_f = 0.1$~au. We choose $\rho_0 = 10^{-9}~\mathrm{g\ cm^{-3}}$, $r_0 = 6.3\times10^{-2}~\mathrm{au}$, and $p = -1.35$, which yields a total mass in the 10~au simulation domain of $10^{-3}~M_\odot$. 

Following \citet{Flock2016}, we set the temperature in the disk (where $\vert z/r \vert< 0.1$) as 
\begin{equation}
    T_\mathrm{d}(r) = \frac{T_\odot}{\epsilon_T^{0.25}}\left(\frac{R_\odot}{2r}\right)^{1/2}
\end{equation}
where $T_\odot=5666$K and $R_\odot = 6.943\times10^{10}$cm are the surface temperature and radius of the Sun, respectively, and $\epsilon_T = 1/3$. Above the disk surface, we assume a temperature transition region up to an angular height $\vert z/r\vert = 0.3$. The temperature above the transition region (where $\vert z/r\vert>0.3$) is assumed to be uniform $T_\mathrm{env} = 3,000$~K. The temperature in the transition region is given by
\begin{equation}
    T_\mathrm{trans}(r) = (T_\mathrm{env} - T_\mathrm{d})\frac{\vert z/r\vert - 0.1}{0.2} + T_\mathrm{d}; \ \ 0.1\leq \Big\vert\frac{z}{r}\Big\vert \leq 0.3
\end{equation}
to allow a smooth transition between disk temperature and envelope temperature. To better simulate the active zone and account for accretion heating, we artificially increase the temperature within a radius of 0.1~au and between 0.1 and 0.2~au by, respectively,
\begin{equation}
    T_\mathrm{raised}(r)_{r < 0.1\mathrm{au}} = 2~T(r)
\end{equation}
and
\begin{equation}
    T_\mathrm{raised}(r)_{0.1\mathrm{au} \leq r < 0.2\mathrm{au}} = \Big(3 - \frac{r}{0.1\mathrm{au}}\Big)T(r)
\end{equation}
This temperature profile gives an active zone spanning from the inner boundary at $r=0.03$~au to about $0.13$~au. We assume the spatial temperature profile remains unchanged throughout the simulation to simplify the calculation and increase numerical stability.

The initial magnetic field is calculated by taking the curl of a vector potential to ensure a divergence-free magnetic field. Following \citet{Wang2019}, the vector potential is given by $\mathbfit{A} = \langle 0, 0, A_\phi \rangle $, where
\begin{equation}
    A_\phi = \frac{4}{3} \frac{B_{z, 0}(r/r_0)^{-0.25}}{[1+1/(\mu\tan\theta)^2]^{0.625}};
\end{equation}
$\mu = 0.5$ determines how curved the initial field is. We set the scale for the magnetic field strength $B_{z, 0} = 9\times 10^{-3}~\mathrm{G}$ so the plasma-$\beta$ at $1~\mathrm{au}$ is about $10^5$. 
We note that this vector potential yields a purely poloidal field that initially threads the disk and opens to infinity. It is used to parameterize the intrinsic disk field \citep[inherited, e.g., from the magnetized star-forming molecular cloud cores,][]{Tu2024a}, rather than that from the stellar magnetosphere since our goal is to study the limiting case of purely disk-driven outflow in the absence of a stellar magnetosphere. {The initial disk field is about 1~G at the inner edge of the disk at a radius of 0.03~au, which is weaker than the typical stellar magnetospheric field expected at a similar radius. Nevertheless, it is quickly amplified by differential rotation, leading to a much stronger toroidal field at later times (Sec.~\ref{subsec:overview}). A stronger initial poloidal field may launch a faster outflow, everything else being equal.}

In the radial direction, the hydro boundary conditions for both the inner and the outer boundary is the ``no inflow'' boundary: if the gas flows out of the simulation domain, the gas velocity at the boundary is copied into the ghost zone; if the gas flows into the simulation domain, the gas velocity is set to 0 in the ghost zones. The magnetic boundary conditions at both radial boundaries are the ``standard'' zero-gradient boundary conditions, where the value at the first active cell is copied into the ghost zones. This is a conservative boundary condition to ensure that only the outflow driven by the processes inside the computation domain is captured in the simulation; the outflow could be stronger than what we simulate here if the poloidal field lines passing through the inner radial boundary are attached to rapidly rotating material inside the boundary and constantly twisted by it. Reflective boundary conditions are used in the polar direction.

\subsection{Magnetic diffusivity}
Ohmic dissipation is included in our model since it is important in the relatively cool midplane region of the disk (roughly beyond 0.13~au in our model) where the gas is insufficiently ionized thermally or by (attenuated) cosmic rays \citep[e.g.][]{Gammie1996}. 
The Ohmic diffusivity is calculated by combining the contributions of two ionization sources: thermal ionization and cosmic ray ionization. It is given by 
\begin{equation}
    \eta_O = \frac{c^2}{4\pi\sigma_O},
\end{equation}
where $c$ is the speed of light and the conductivity $\sigma_O$ is dominated by electrons 
\begin{equation}
    \sigma_{O} \approx \frac{n_e e^2}{m_e\gamma_s\rho}
\end{equation}
where $n_e$ and $m_e$ are the electron number density and mass, and $\gamma_s$ is the momentum transfer rate between electrons and neutrals. 

In the innermost part of the disk, where the temperature exceeds $\sim 10^3$~K, thermal ionization dominates. We compute the electron number density $n_e$ from the Saha equation, including four elements with relatively low ionization potential $e_I$ (see Table~\ref{tab:ions}). In the cooler region at larger radii, the ionization fraction drops rapidly, leading to a dead zone with a very large Ohmic diffusivity. To prevent the time scale for the magnetic diffusion from becoming prohibitively small, we include cosmic ray ionization using the simple prescription of \cite{Shu1992}:
\begin{equation}
    n_e = n_i = \frac{C\sqrt{\rho}}{m_i},
\end{equation}
where $n_i$ is the number density of the dominant molecular ion and $m_i \approx 29\ m_H$. $\rho$ is the local gas density. We choose $C = 9.5\times 10^{-18}~{\rm cm}^{-3/2}~{\rm g}^{1/2}$, corresponding to a cosmic ray ionization rate $10^3$ times lower than the canonical value of $10^{-17}$~s$^{-1}$ to account for the cosmic ray attenuation expected in the dead zone; it is also comparable to the ionization rate expected from radioactive nuclides \citep[][]{Umebayashi2009}. 
The better magnetic coupling in the surface layer and envelope of the disk (which is more ionized by the less attenuated cosmic rays and energetic photons such as UV and X-rays) is represented by an increased temperature and its associated thermal ionization.


Following \citet[][]{Wang2019}, we include an inner buffer zone from a radius $r=0.045$~au to the inner boundary at $r=0.03$~au to increase numerical stability, where the magnetic diffusivity is gradually decreased from its original value (at 0.045~au) to 0 (at 0.03~au). Since the magnetic field is already well coupled to the gas due to a relatively high temperature (and associated efficient ionization of alkali metals), we do not expect the further reduction in the magnetic diffusivity to affect the results significantly. { We do not include any anomalous magnetic diffusivity (due to, e.g., sub-grid turbulence) in the simulation, which may affect the magnetic field evolution and gas dynamics in magnetically well-coupled regions that are MRI active, although some effective turbulent magnetic diffusion due to reconnection is captured in our simulation (see Section \ref{subsec:reconnection}).}

\begin{table}
    \centering
    \begin{tabular}{c c c c}
         Species & $m_s$ [$m_p$] & $n/n_\mathrm{H2}$ & $e_I$ [erg]\\
         \hline
         K & $39$ & $2.14\times10^{-9}$ & $6.95\times10^{-12}$ \\
         Ca & $40$ & $4.38\times10^{-8}$ & $9.76\times10^{-12}$ \\
         Na & $23$ & $3.47\times 10^{-8}$ & $8.2\times 10^{-12}$ \\
         Mg & $24$ & $7.96\times 10^{-8}$ & $1.2\times 10^{-11}$
    \end{tabular}
    \caption{Atomic species included in the thermal ionization calculation using the Saha equation. $m_s$ in the second column is the atomic mass of each element in the unit of proton mass $m_p$. The abundances in the third column are taken from \citet{Asplund2009}, assuming 1\% of the metal is in the gas phase. The last column is the ionization potential $e_I$.}
    \label{tab:ions}
\end{table}

\section{Simulation Results}
\label{sec:result}

\subsection{Overview}
\label{subsec:overview}


\begin{figure}
    \centering
    \includegraphics[width=\columnwidth]{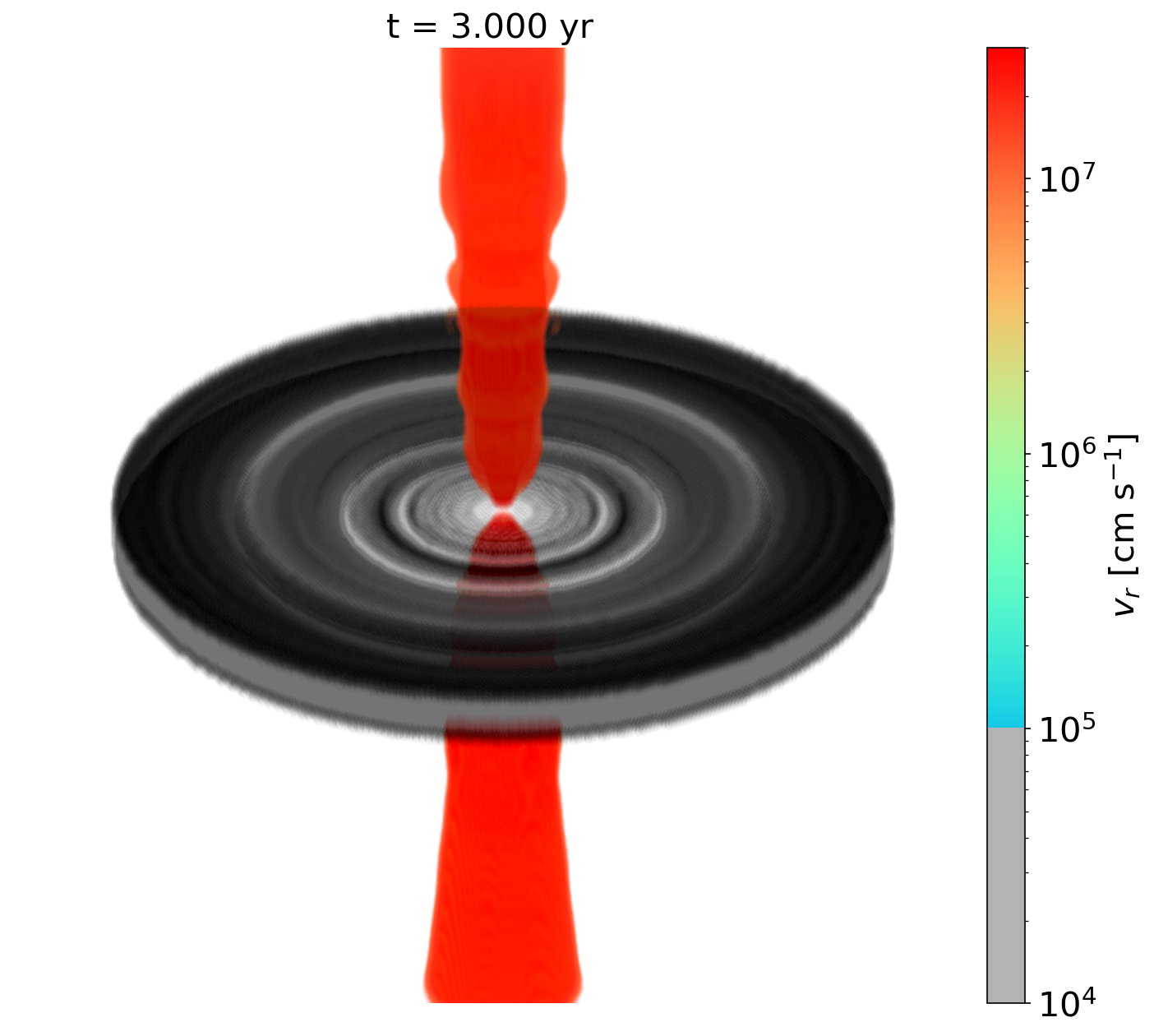}
    \caption{3D rendition of the disk-jet system at a representative time $t=3.0$~yr, showing a fast ($>100$~km/s) collimated bipolar jet (colored red) driven from a highly dynamic dense equatorial disk (colored purple and blue). The jet maintains its fast speed 
    reaching the upper and lower edges of this rendition, which are at the boundaries of the simulation domain ($10$~au in the $\hat{z}$ direction). The horizontal black structure is a representation of the disk up to $5$~au in radius; all features in the disk representation are rendering effects. An animated version of this figure is available in the online journal. The movie is 9 seconds long, showcasing a continuous jet launched in our model.}
    \label{fig:3D}
\end{figure}

\begin{figure*}
    \centering
    \includegraphics[width=\textwidth]{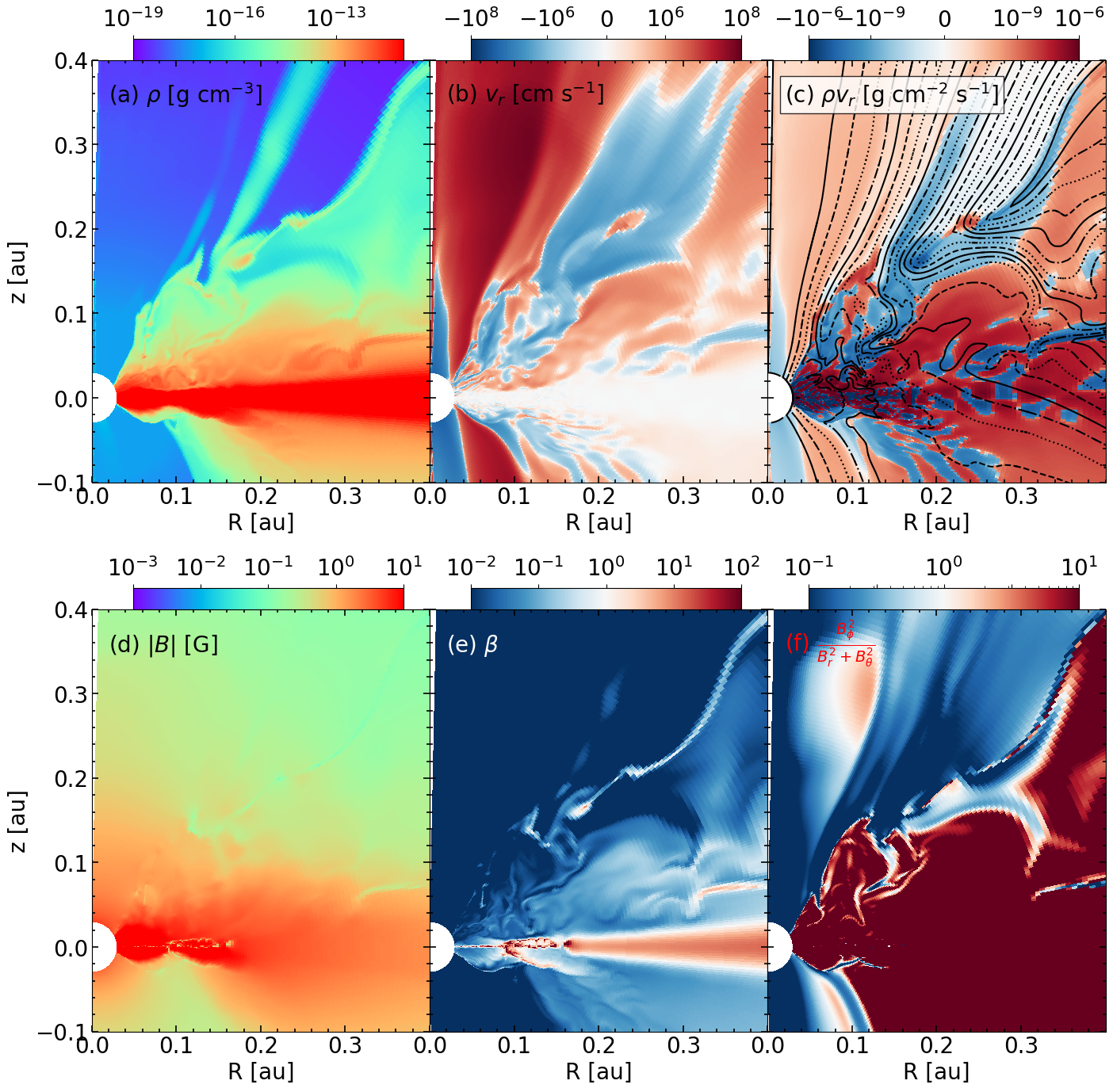}
    \caption{Overview of the simulation result on a meridional plane at a representative time $t=3.0$~yr. The upper panels show (a) the density, (b) radial velocity $v_r$, and (c) the mass flux, respectively. The lower panels show (d) the magnetic field strength, (e) plasma-$\beta$, and (f) the ratio between the toroidal field and the poloidal field, respectively. An animated version of this figure is available in the online journal. The movie is 27 seconds long, showing an overview of the disk and outflow evolution in our model.}
    \label{fig:overview}
\end{figure*}

To give a first impression of the disk-jet system, we show in Fig.~\ref{fig:3D} a representative frame of an animation of the simulation. It shows clearly that a fast ($>100$~km/s) collimated bipolar jet (colored red)\footnote{ The minimum (threshold) speed of 100~km/s for highlighting the jet in the animation is chosen somewhat arbitrarily; it is on the low side of the observed range of YSO jet speed, especially for the atomic jets from relatively evolved Class II sources. It does not mean that all jet materials move at 100 km/s; they have a range of speed up to at least 500~km/s.} is continuously driven along the north and south poles of the dense circumstellar disk (colored black and gray). Fig.~\ref{fig:overview} gives a more detailed view of the distributions of various flow quantities in the inner part of the simulation, focusing on the upper hemisphere. 

The top row of Fig.~\ref{fig:overview} shows the density (panel a), radial direction velocity $v_r$ (panel b), and the mass flux $\rho v_r$ (panel c) at a representative time $t=3.0$~yr, corresponding to about 577 times the orbital period at the inner disk edge. 
The poloidal magnetic field lines, with an equally spaced poloidal magnetic flux between adjacent field lines, are overplotted on the mass flux panel. 
The disk resides near the equatorial plane where $z=0$, highlighted by the high density and low $v_r$. It is surrounded by a more diffuse and faster-moving atmosphere.    
Close to the polar axis is an outflow (jet), where the density is lower, and flow is faster, reaching values as high as $\sim 10^8$~cm/s. The poloidal magnetic field lines are concentrated in the jet region, suggesting a close relationship between the outflow and the magnetic field.

The magnetic field of the simulation is quantified in the lower panels of Fig.~\ref{fig:overview}, which display the magnetic field strength (panel d), plasma-$\beta$ (panel e), and the ratio between toroidal magnetic pressure and poloidal magnetic pressure ($B_\phi^2 / (B_\theta^2 + B_r^2)$ (panel f). The strongest magnetic field region in panel (d) is the active zone on the disk (extending to about 0.13 au in radius), where the temperature is high enough to ionize metals with low ionization potentials. The gas is close to the ideal MHD limit in this region. As a result, the magnetic field builds up through differential rotation in the active zone, reaching a strength $>10$G. Consequently, the plasma-$\beta$ in the active zone stays mostly of order or below unity. Outside the active zone in the disk is the dead zone, where the temperature near the midplane is insufficient to ionize the metals. Ohmic dissipation is strong enough to prevent rapid amplification of the magnetic field, resulting in a plasma-$\beta$ greater than 1. The gas is well ionized in the disk atmosphere so that it is close to the ideal MHD limit. Fig.~\ref{fig:overview}f shows that the poloidal field dominates the polar region (the solid blue region close to the poles), whereas the toroidal field dominates the disk and part of the disk atmosphere (the solid red regions). Between these two extremes is a region where the toroidal and poloidal field components are more comparable. The fast jet resides in this region (Fig.~\ref{fig:overview}b). 

\begin{figure}
    \centering
    \includegraphics[width=\columnwidth]{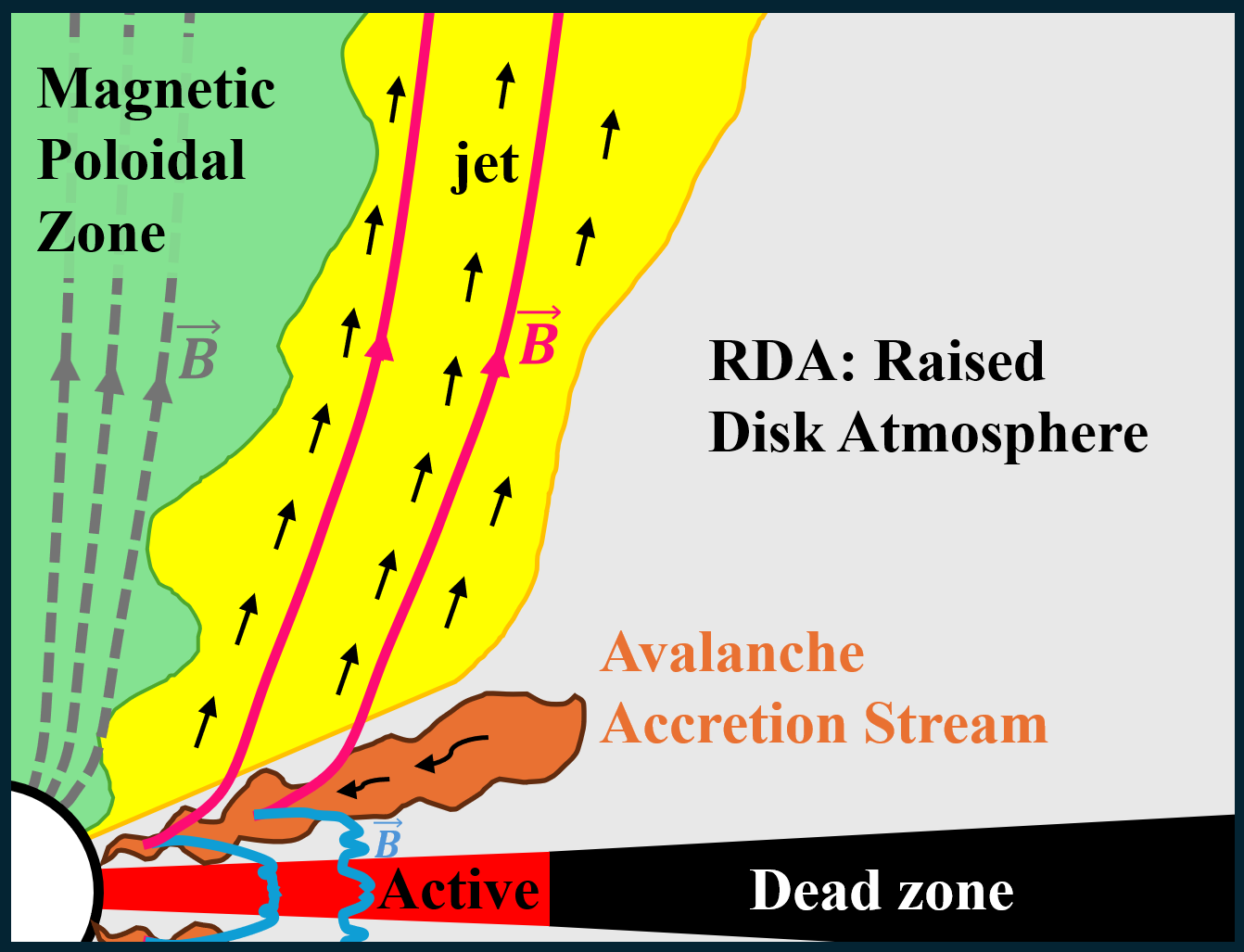}
    \caption{Key components of the avalanche accretion stream driven jet-launching model. The green zone is the low-density magnetic poloidal zone (with the dominant poloidal field highlighted by gray dashed lines with arrows), surrounded by the jet. The gray zone is the raised disk atmosphere (discussed in sec.~\ref{subsec:atmosphere}), and the orange stream is the avalanche accretion stream (discussed in sec.~\ref{subsec:AvalancheStreams}). The disk is divided into the active (red) and the dead zone (black). The cyan and magenta line segments highlight the magnetic field geometry, with a sharp pinch on the avalanche accretion stream (discussed in sec.~\ref{subsec:outflow_launching}).  The black arrows highlight the gas motion in the avalanche accretion stream and the jet.} 
    \label{fig:zone_sketch}
\end{figure}

To facilitate the discussion of the jet launching mechanism, we conceptually divide the simulation domain into several zones based on their physical characteristics. The zoning is sketched in Fig.~\ref{fig:zone_sketch}. The disk is divided into an inner active and outer dead zone based on their temperatures. Above the disk is the avalanche accretion stream, which drags the poloidal magnetic fields (the magenta and cyan lines) into a sharply pinched configuration. A jet is launched above the avalanche accretion stream along the upper (magenta-colored) branch of the pinched field line (the yellow region, with arrows showing the gas flow direction). 
The avalanche accretion streams are embedded in a thick atmosphere (colored gray in the sketch), which surrounds the jet region in the cylindrically radial direction and helps confine it laterally.  Since the jet-launching avalanche streams reside in the raised atmosphere, it will be discussed next.


\begin{figure*}
    \centering
    \includegraphics[width=\textwidth]{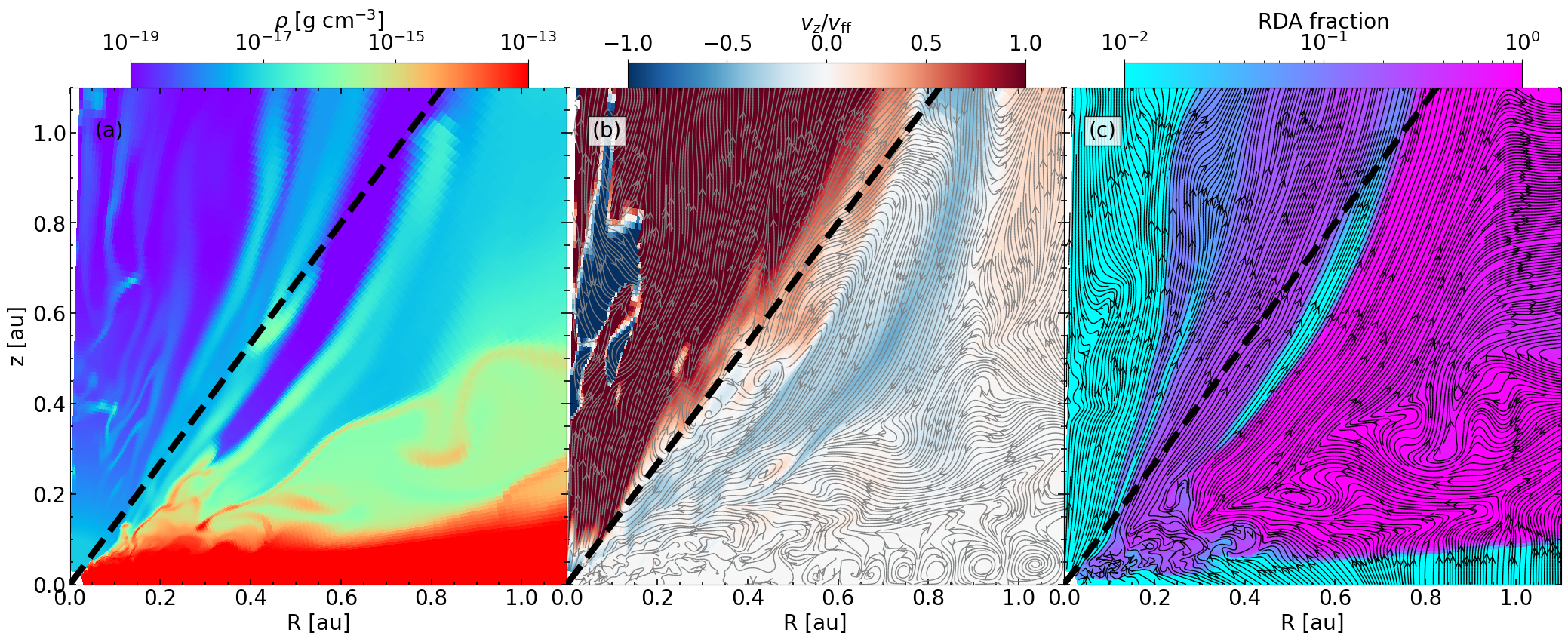}
    \caption{The properties of the raised disk atmosphere (RDA), roughly located to the right of the dashed black line, at a representative time of $t=3.2$~yr. Plotted are (a) the density distribution, (b) the vertical gas velocity normalized by the local free-fall speed, and (c) the fraction of mass coming from the material originally near the disk surface (see text). Overplotted in panel (b) are the poloidal velocity streamlines, and in panel (c) are the poloidal magnetic field lines. }
    \label{fig:MASA6p}
\end{figure*}

\subsection{Raised Disk Atmosphere}
\label{subsec:atmosphere}

Fig.~\ref{fig:MASA6p}a and b show the density and vertical velocity component normalized by the local free-fall speed $v_z/v_\mathrm{ff}$ of the raised disk atmosphere (RDA), respectively. The RDA is the clumpy region above the disk with a relatively high density (compared to the more evacuated polar region) 
but a relatively low vertical velocity $v_z$ compared to the local free-fall (or escape) speed, indicating that it remains bound to the disk through stellar gravity. 
Although some parts of the RDA move away from the disk, the motion is too slow to escape from the star's gravity. 
Thus, distinct from the much faster-moving jet in the polar region, this RDA is essentially a quasi-static structure enveloping the disk, i.e., a raised disk atmosphere. Above the active disk zone and the inner dead zone, we find that the magnetic pressure dominates the thermal pressure, with the magnetic pressure gradient from the toroidal field primarily responsible for raising the atmosphere. At larger radii, the thermal pressure gradient becomes relatively more significant, although it takes longer for the toroidal field to grow through the differential twisting of the initial poloidal field at larger distances.  

To illustrate more pictorially that the vertically raised atmosphere was part of the material originally closer to the disk, we mark the material initially in the disk surface region between $0.1 < \vert z/r\vert < 0.15$ and $r > 0.2~\mathrm{au}$ with a scalar of 1. Fig.~\ref{fig:MASA6p}c plots the spatial distribution of the scalar value at the representative time, showing that the material originally close to the disk surface is now raised to occupy most of the atmosphere (the pink region in the panel). The poloidal magnetic field lines are overplotted in the panel, showing pinched morphologies in several regions of the atmosphere, particularly in the lower-left quadrant of the panel. These pinched magnetic field lines characterize the avalanche accretion streams, which we will focus on in the following subsection.

\subsection{Avalanche Accretion Streams}
\label{subsec:AvalancheStreams}

\begin{figure*}
    \centering
    \includegraphics[width=\textwidth]{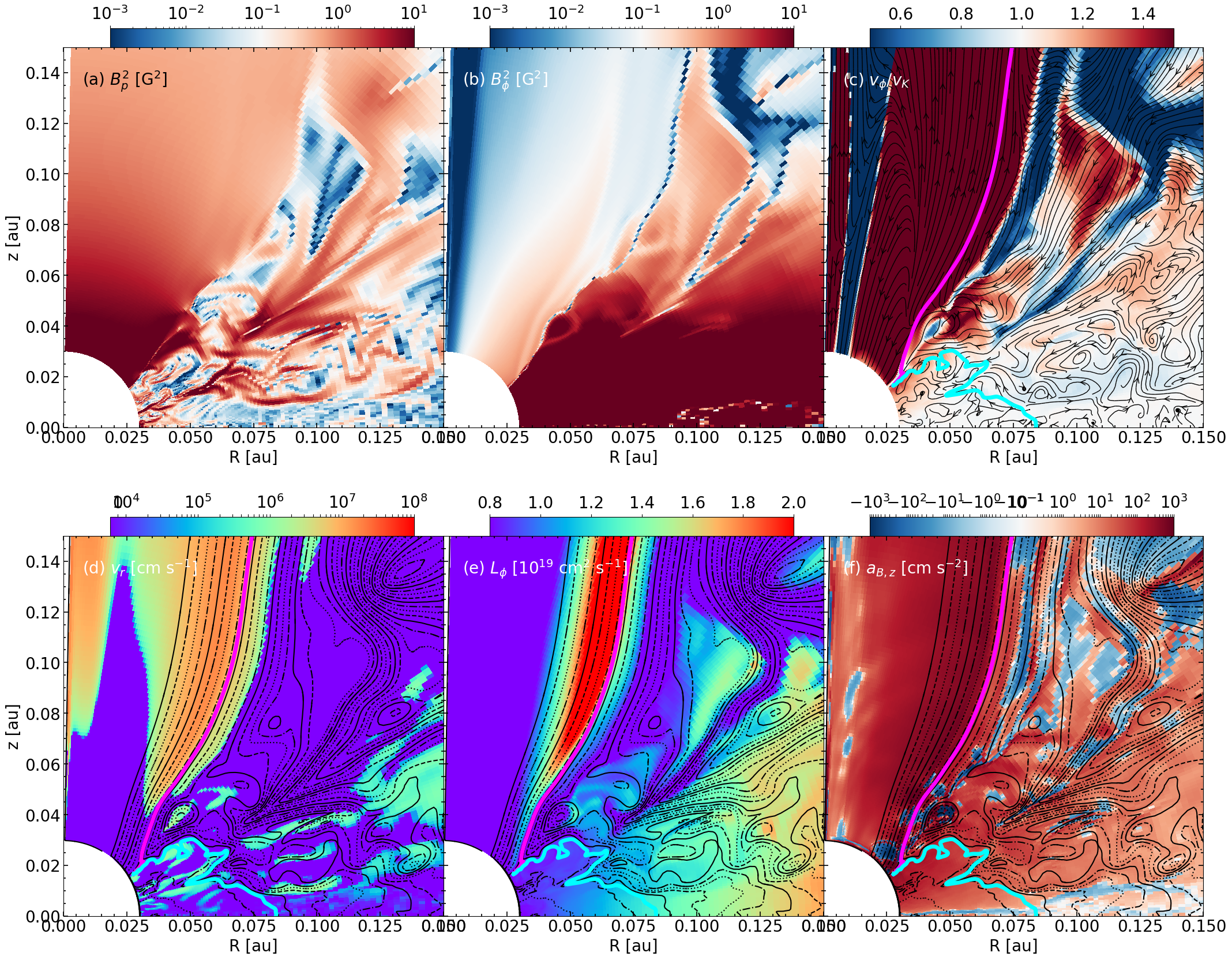}
    \caption{The disk and outflow properties at a representative time of $t=3.01$~yr when a prominent avalanche accretion stream is present. Panels (a) and (b) show the squared poloidal and toroidal magnetic field strength, respectively. Panel (c) shows the ratio of the toroidal velocity and local Keplerian velocity,  with the poloidal velocity streamlines overplotted. The lower panels show (d) the radial gas velocity $v_r$, (e) the specific angular momentum, and (f) the vertical magnetic acceleration, respectively. The poloidal magnetic field lines on the lower panels are plotted as contours of equal radial magnetic flux. {A representative poloidal field line threading the avalanche accretion stream is highlighted in panels (c)-(f), with its pinch tip dividing the highlighted field line into two branches. The upper branch is colored magenta, and the lower branch is colored cyan. Flow acceleration along this representative field line is discussed in Section~\ref{subsec:outflow_launching} below.} 
    }
    \label{fig:avalanche_3panel}
\end{figure*}

We now switch our focus to smaller cylindrical radii, where the magnetic poloidal zone, jet zone, active disk zone, and prominent avalanche accretion streams reside. Since the avalanche accretion streams are crucial to jet-launching (sec.~\ref{subsec:outflow_launching}), we describe their properties and formation in this subsection.

The avalanche accretion stream is sandwiched between the active disk zone and the magnetic poloidal zone, where the poloidal magnetic flux accumulates through mass accretion. At the time shown in Fig.~\ref{fig:avalanche_3panel}, 
the poloidal magnetic flux threading the inner (spherical) boundary in the region from $\theta=0$ to $\theta=\pi/4$ increases by about a factor of $\sim 6$ compared to the initial flux. The magnetic field strength in the active disk zone is amplified when the poloidal field is winded into the toroidal field. Because the active zone is close to the ideal MHD limit, non-ideal MHD effects cannot remove the generated toroidal flux. As a result, a strong toroidal field develops in the active zone. To show the magnetic properties of these two zones more vividly, we show in Fig.~\ref{fig:avalanche_3panel}a and b the square of the poloidal and toroidal field strength, respectively. The magnetic poloidal zone is dominated by a strong poloidal field (see Fig.~\ref{fig:avalanche_3panel}a), while the active disk zone and the atmosphere immediately above it are dominated by a strong toroidal field (see Fig.~\ref{fig:avalanche_3panel}b).

These two zones occupy most of the volume and angular space around the inner (radial) boundary. To accrete the raised atmosphere through the inner boundary, both zones must be avoided along the trajectory of an accreting gas parcel. As a result, the gas can only travel through a narrow corridor between these two zones, in the form of narrow streams. In order for the streams to accrete rapidly, they must be braked strongly. 

Fig.~\ref{fig:avalanche_3panel}c shows the avalanche accretion stream in the narrow corridor is braked so strongly that it rotates at a significantly sub-Keplerian speed. 
The sub-Keplerian nature of the accretion flow results from a positive feedback loop between magnetic braking and accretion, where angular momentum removal by magnetic braking enables the gas near the tip of the pinched poloidal field to sink closer to the central star, where it rotates faster and is magnetically braked more strongly, leading to an avalanche-like gas falling, which is a form of the magneto-rotational instability \citep[e.g.][]{Kudoh1998, Suriano2017, Zhu2018, Mishra2020}. 
Specifically, a gas parcel moving radially inwards along the narrow corridor pinches the poloidal magnetic field line on a meridional ($R-z$) plane (see Fig. \ref{fig:avalanche_3panel}c) and the toroidal magnetic field line on the cylindrical $\phi-z$ plane. The pinch on the cylindrical $\phi-z$ plane brakes the gas rotation magnetically. We can divide the poloidal magnetic field line from the tip of the pinch into two branches, as illustrated by the highlighted fieldline in Fig.~\ref{fig:avalanche_3panel}c. The upper branch is colored magenta, and the lower is colored cyan. Because the upper branch is restricted radially by the magnetic poloidal zone and the lower one is embedded in the heavily mass-loaded active disk zone, neither branch of the magnetic field line can move inwards together with the tip of the pinch. As a result, when the gas accretes further, the poloidal field line becomes more pinched, which, in turn, brakes the gas harder, facilitating further infall. This feedback loop creates the avalanche accretion stream, which is responsible for fast outflow launching, as we demonstrate next. 

\subsection{Magnetocentrifugal Jet Driven by Avalanche Accretion Streams}
\label{subsec:outflow_launching}

\begin{figure*}
    \centering
    \includegraphics[width=\textwidth]{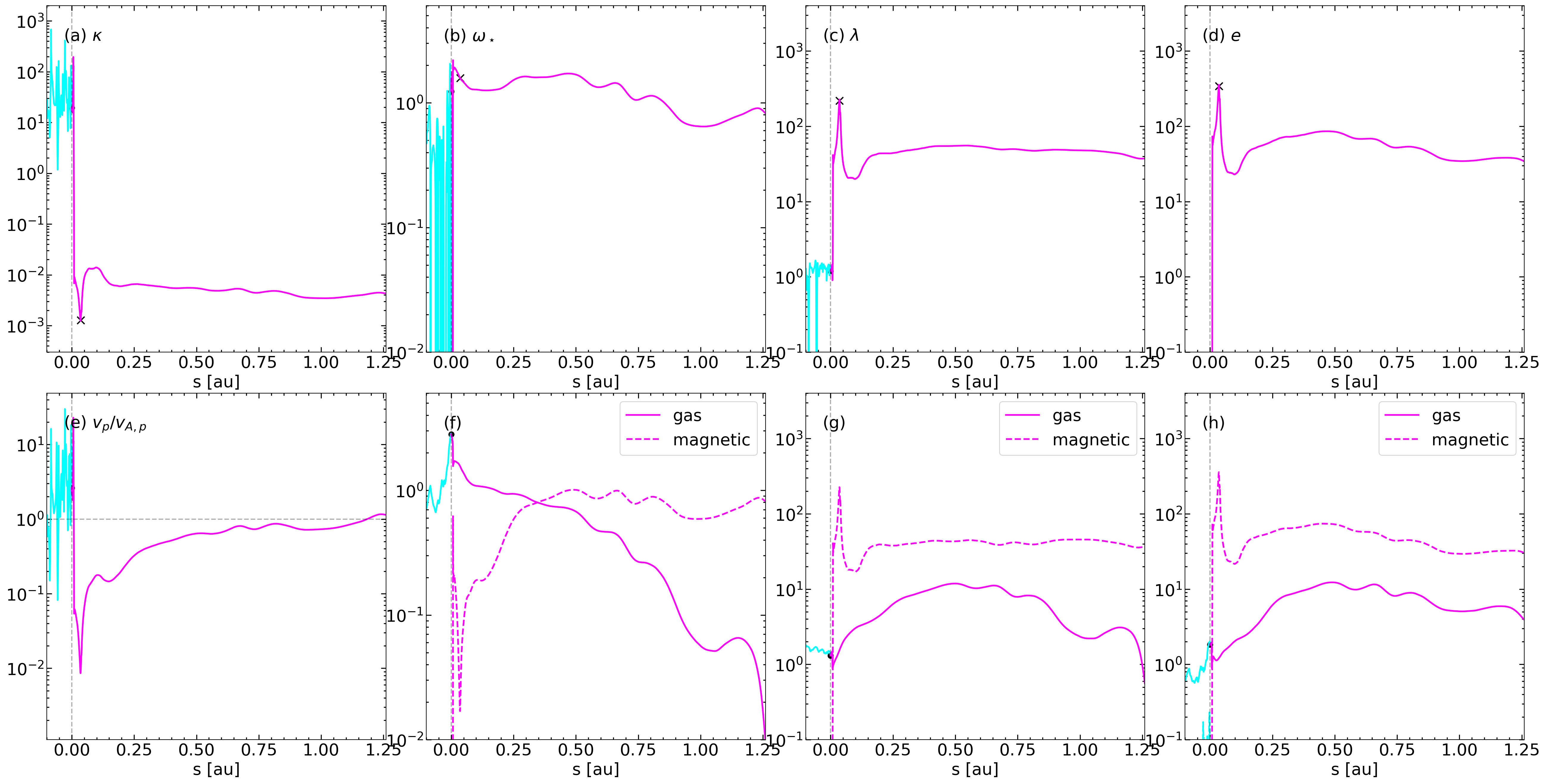}
    \caption{Quantities along the highlighted magnetic field line in Fig.~\ref{fig:avalanche_3panel}. $s = 0$ on the horizontal axis corresponds to the tip of the pinched field line; the $s > 0$ section follows the upper branch of the field line (along the magenta field line), and the $s < 0$ section follows the lower branch (along the cyan field line). The distances shown are projected along the field line on the meridian plane. The upper panels are the four conserved quantities defined in equ.~\ref{equ:cons:k}, \ref{equ:cons:O}, \ref{equ:cons:l}, \ref{equ:cons:e} respectively. Panel (e) shows the ratio between the magnitude of the poloidal velocity and the poloidal Alfvén wave speed. The gas speed becomes super-Alfvénic at $\sim 1.2~\mathrm{au}$ along the magenta field line. Panel (f), (g), and (h) show the gas and magnetic components of equ.~\ref{equ:cons_omega}, \ref{equ:cons_angmom}, and \ref{equ:cons_ene} respectively. The gas components are the first terms in equ.~\ref{equ:cons_omega}, \ref{equ:cons_angmom}, and \ref{equ:cons_ene}; the magnetic components are the second terms in equ.~\ref{equ:cons_omega}, \ref{equ:cons_angmom}, and the third term in equ.~\ref{equ:cons_ene}. 
    }
    \label{fig:fieldline_cons}
\end{figure*}

From Fig.~\ref{fig:overview} and its associated animation, we observe that fast outflow tends to be driven along the upper branch of the highly pinched poloidal magnetic field of an avalanche accretion stream that extends to a large radius in a relatively evacuated region, such as the magenta line in Fig.~\ref{fig:avalanche_3panel}. The outflow is generated by the conversion of magnetic energy into gas kinetic energy, as in a standard MHD wind \citep[see, e.g.,][for a review]{Spruit2006}.
Although the outflow is highly time variable, it is still instructive to analyze the four quantities that are conserved along a magnetic field line for a steady axisymmetric wind \citep{Weber1967, Blandford1982}: the mass loading parameter
\begin{equation}
    k = \frac{4\pi\rho v_p}{B_p};
\end{equation}
the angular speed 
\begin{equation}
    \omega_s = \frac{v_\phi}{R} - \frac{k B_\phi}{4\pi \rho R};
    \label{equ:cons_omega}
\end{equation}
the specific angular momentum
\begin{equation}
    l = R v_\phi - \frac{RB_\phi}{k};
    \label{equ:cons_angmom}
\end{equation}
and the specific energy
\begin{equation}
    E = \frac{1}{2}v^2 + \Phi - \frac{\omega_s R B_\phi}{k} + h,
    \label{equ:cons_ene}
\end{equation}
where $R$ is the cylindrical radius, $v_p^2 = v_r^2 + v_\theta^2$, $B_p^2 = B_r^2 + B_\theta^2$, and $h$ is the enthalpy per unit mass. These dimensional quantities can be non-dimensionalized using the gas properties at the base of the outflow, which is taken to be the slow magnetosonic point \citep[denoted with a subscript ``SM,"][]{Jacquemin-Ide2021, Blandford1982}
\begin{equation}
    \kappa = k \frac{v_\mathrm{K, SM}}{B_\mathrm{z, SM}}
    \label{equ:cons:k}
\end{equation}
\begin{equation}
    \omega_\star = \frac{\omega_s}{\Omega_\mathrm{K, SM}};
    \label{equ:cons:O}
\end{equation}
\begin{equation}
    \lambda = \frac{l}{R_\mathrm{SM} v_\mathrm{K, SM}};
    \label{equ:cons:l}
\end{equation}
\begin{equation}
    e = \frac{E}{v^2_\mathrm{K,SM}},
    \label{equ:cons:e}
\end{equation}
where the subscript ``K'' denotes the Keplerian value, with $v_\mathrm{K, SM}=R_\mathrm{SM}\Omega_\mathrm{K, SM}$. Note that the normalization for the specific energy, $v^2_\mathrm{K,SM}$, is consistent with that in \cite{Blandford1982}, which is a factor of 2 larger than that used in \cite{Jacquemin-Ide2021}.

Fig.~\ref{fig:fieldline_cons} shows these four dimensionless conserved quantities on the magnetic field line highlighted in the lower panels of Fig.~\ref{fig:avalanche_3panel} and the contributions by the fluid and magnetic field to three of the conserved quantities on the last three lower panels. The dimensionless mass loading parameter $\kappa$ (Fig.~\ref{fig:fieldline_cons}a) is around $30$ along the lower branch (colored cyan) of the magnetic field line. The value decreases abruptly to $\sim 10^{-3}$ crossing the tip of the pinched field line from the lower branch to the upper branch (colored magenta) before settling to a value $\sim 4\times 10^{-3}$. The sudden decrease of the mass-loading parameter highlights the transition from the dense magnetically supported quasi-static inner disk atmosphere to the much more tenuous magnetically driven outflow zone. At around 1.2 au along the lightly loaded magnetic field line (see panel e), the gas poloidal velocity exceeds the poloidal Alfv\'en velocity (Fig.~\ref{fig:fieldline_cons}e), defined as $v_{A, p} = \sqrt{\frac{B_p^2}{4\pi\rho}}$ based on the poloidal field strength\footnote{ Note that the flow in the cyan part of the field line in Fig.~\ref{fig:fieldline_cons}e is super-Alfv\'enic with respect to the poloidal magnetic field. The reason is that this part of the field line is located in the much denser part of the disk and disk atmosphere where relatively fast, chaotic poloidal motions are driven by magnetically mediated angular momentum transfer (see Fig.~\ref{fig:avalanche_3panel}c) and the poloidal Alfven speed $v_{A, p}$ is low because of a weak poloidal field. The relatively fast chaotic poloidal motions do not lead to an excessively large mass accretion rate, however, because of the cancellation of inward and outward mass fluxes. Indeed, the average (net) accretion speed (based on the net mass accretion rate) is well below the Alfv\'en speed.}.

The panel (b) of Fig.~\ref{fig:fieldline_cons} shows that the dimensionless quantity $\omega_*$ stays roughly constant outside the slow magnetosonic point (marked by a cross). It is initially dominated by the fluid rotation but becomes increasingly more dominated by the magnetic contribution as the gas angular speed decreases at larger distances. The dimensionless specific angular momentum also approaches an approximately constant value of $\sim 40$ (panel c). The magnetic contribution is more significant than the fluid contribution, but there is a conversion of the former into the latter along the field line (panel g), which gives rise to the super-Keplerian rotation in the jet zone shown in Fig.~\ref{fig:avalanche_3panel}c, which, in turn, facilitates centrifugal outflow acceleration. 

The dimensionless specific energy is also roughly constant in the outflow region (see panel d). It is entirely dominated by the magnetic contribution, even at the largest distance from the base of the outflow shown in the plot (1.2~au), where the magnetic energy remains nearly one order of magnitude higher than the flow kinetic energy (see panel h). Nevertheless, the gas velocity reaches a few$~\times~  10^7~\mathrm{cm/s}$ along the magnetic field line, well above the local escape velocity. 

We note that jet-launching pinched poloidal field lines like the one highlighted in the lower panels of Fig.~\ref{fig:avalanche_3panel} are cut at the inner (radial) boundary when their tips are dragged across the boundary by the accretion streams. A fast outflow persists along the upper branch of the field line immediately after the cut, as the already magneto-centrifugally accelerated material continues to move outward (see, e.g., the dot-dashed and dotted field lines to the immediate right of the magenta line). However, the cut field lines are no longer attached to the dense accreting streams that powered their acceleration before the cut, likely rendering the acceleration captured in the simulation domain a lower limit.

To summarize, a standard magneto-centrifugal outflow appears to be driven from the upper branch of the highly pinched poloidal magnetic field of the avalanche accretion stream in the raised atmosphere of the inner accretion disk. It is collimated into a jet along the rotation axis by the vertically extended, raised disk atmosphere surrounding the outflow, as shown in Fig.~\ref{fig:MASA6p}c. 

\section{Discussion}
\label{sec:discussion}

\subsection{Towards a new picture of protostellar jet formation}
\label{subsec:reconnection}

It is unsurprising that the atmospheric avalanche accretion stream drives the jet. This is because the energy for the jet launching ultimately comes from the release of the gravitational binding energy of mass accretion, which is concentrated in the atmospheric avalanche streams. 

The scenario of the avalanche accretion stream-driven jet is a variant of one of the traditional scenarios of jet launching, where the jet is driven magneto-centrifugally directly from the inner disk \citep[e.g.][the other is the X-wind picture; \citealp{Shu2000}]{Konigl2000, Krasnopolsky2003}. A longstanding concern with the traditional disk-wind picture is that the magneto-centrifugal jet would remove angular momentum from the launching region on the disk so efficiently that it would accrete supersonically, leading to an excessively large accretion rate and rapid depletion of the inner disk as well as accumulation of an excessive poloidal magnetic flux that may cause disk disruption, which may, in turn, choke the jet launching. This ``magnetic flux problem" of jet launching was often alleviated using a prescribed ``turbulent" resistivity that enables the poloidal field to diffuse outward relative to the accreting material \citep[e.g.][]{Stepanovs2016}. 

\begin{figure*}
    \centering
    \includegraphics[width=\textwidth]{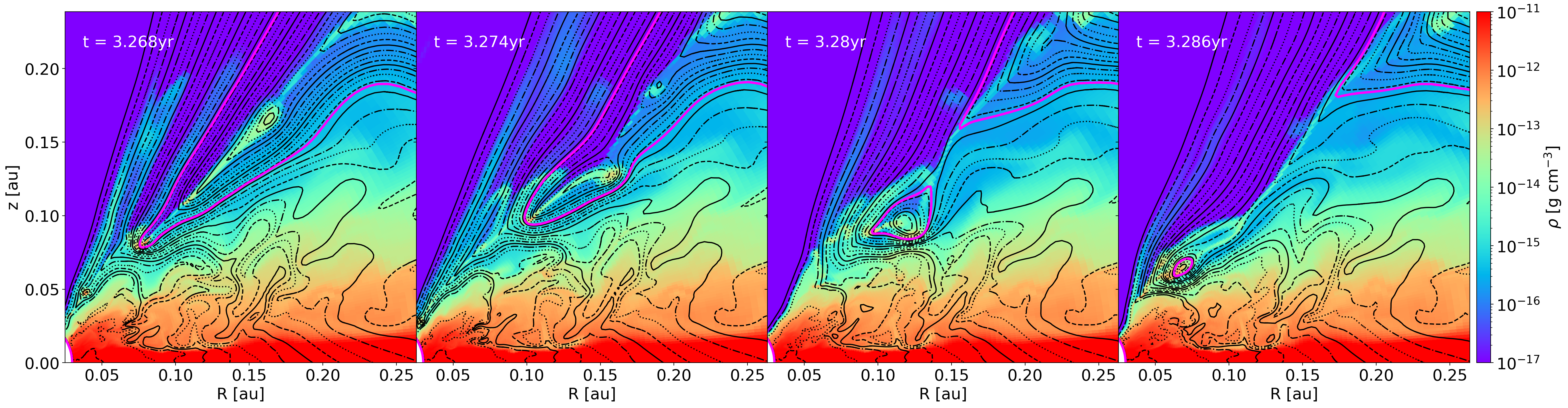}
    \caption{A representative magnetic reconnection event in our model. The four panels are a time sequence of reconnecting magnetic field lines. The magnetic field lines are specified by their values of the poloidal magnetic flux so that individual field lines can be traced over time. The magenta field line highlights a specific reconnecting magnetic field line. An animated version of this figure is available in the online journal. The movie is 27 seconds long, which highlights the magnetic reconnection along the avalanche accretion stream. The highlighted magenta field line reconnected multiple times during the movie.}
    \label{fig:reconnection2D}
\end{figure*}

Rapid inflow is also present in the atmospheric avalanche accretion stream in our picture, but it does not lead to a mass accretion rate as large nor a poloidal magnetic flux accumulation as fast as in the traditional picture. The former is because the rapid inflow involves only a small fraction of the raised disk atmosphere, which is much less dense than the disk midplane. The latter is because the highly pinched poloidal field lines in the accretion streams are prone to reconnection, allowing the mass to accrete without dragging along the field lines. 


In Fig.~\ref{fig:reconnection2D}, we show an example of the poloidal field line evolution involving reconnection. The lines shown are the contour lines of the poloidal magnetic flux, with each field line specified by a distinct flux value; for example, the magenta line in different panels marks the same field line at different times. From the time sequence, it is clear that the highlighted field line reconnected. For example, the highly pinched field line in the first and second panels has reconnected in the third and fourth panels, with the detached magnetic loop falling inward and the remaining reconnected open field line moving outward.  This is a clear demonstration that accreting matter with a closed magnetic loop does not lead to the accretion of magnetic flux (open field line), which is left behind. The same process repeats itself, as shown in the animated version of Fig.~\ref{fig:reconnection2D} (see the figure caption for a link to the animation). 

The reconnecting atmospheric avalanche accretion streams that play an essential role in mass accretion and outflow launching correspond to the so-called ``turbulent atmosphere" defined by \citet{Jacquemin-Ide2021} in their 3D simulations (marked as ``TA" in their Fig.~3, top panel). The ridges of these streams have weaker magnetic fields than their surroundings because of the sharp pinching of field lines in both poloidal and toroidal directions. This leads to vanishing field components along the accretion stream and in the toroidal direction at the locations where the field lines reverse direction. We believe this is why the time-averaged field strength in their ``turbulent" layer has a dip (see the red line in the shaded regions near the polar angle $\theta=\pi/4$ and $3\pi/4$ in their Fig.~7), which makes it easier for the time-averaged component of the field strength to drop below the time-varying (fluctuating) component (see the blue dashed line of the figure), the defining characteristic of their turbulent atmosphere. The correspondence can be seen even more explicitly from the top panel of their Fig.~8, which shows that their turbulent atmosphere is near the region where their time-averaged poloidal field lines are sharply pinched, as in our (axisymmetric) avalanche accretion streams. 

One advantage of our 2D (axisymmetric) simulations is that each poloidal field line can be labeled by the poloidal magnetic flux it encloses, which enables us to show explicitly the reconnection of the field lines in the avalanche accretion streams (see Fig.~\ref{fig:reconnection2D} and especially its animated version). The reconnection is crucial for the outward transport of the poloidal magnetic flux related to the accreted gas, which was attributed to a turbulent ``effective" resistivity by \citet{Jacquemin-Ide2021}. 
We conclude that the fast outflows in our and their simulations are both driven by atmospheric avalanche accretion streams, but the results are interpreted from different perspectives. 

Atmospheric avalanche accretion streams have been observed in other simulations in the context of both AGN (Active Galactic Nucleus) and YSO (Young Stellar Object) disk accretion, particularly when the disk is initially threaded with a relatively weak large-scale open poloidal field. For example, \citet{Matsumoto1996} carried out 2D (axisymmetric) ideal MHD simulations of the accretion of an initially thick toroid threaded by a uniform poloidal field onto a central black hole. They found a fast outflow driven by magnetically braked surface accretion streams, which they first termed ``avalanche accretion." They interpreted the outflow launching as a consequence of the outward propagation of non-linear Alfv\'en waves due to twisted field lines. Despite the differences in our simulation setup and somewhat different interpretations of the outflow production (magneto-centrifugal launching vs Alfv\'en waves), the jet production mechanism explored in this paper can be viewed as the YSO counterpart to their mechanism. Similarly, \citet{Beckwith2009} simulated the accretion of a magnetized toroid onto a central black hole in 3D, finding high-latitude accretion streams with highly pinched poloidal field lines (which they called "hairpins") and evidence for reconnection (as in our case). However, their focus was on magnetic flux transport rather than outflow generation. 

In the context of YSO disk accretion, \citet[][]{Takasao2018} appears to contradict our results because it has prominent accretion streams but no obvious fast outflow. Their accretion streams are located near the outer boundary enclosing a low-density polar region filled with a strong, ordered, poloidal magnetic field referred to as a ``funnel." They are termed ``funnel-wall accretion," which plays a crucial role in their picture of mass accretion onto a non-magnetized central star. These funnel-wall accretion streams are produced by strong magnetic braking, just like our avalanche flows, but they do not appear to drive an easily discernable fast outflow. It is not exactly clear why a fast outflow is driven magnetically by the accretion streams in our case but not theirs. One possibility is that the magnetically driven fast outflow may be masked by the powerful thermally driven stellar wind included in their case; the absence of a stellar wind may have allowed the accretion stream-driven fast outflow to show up more clearly in our case. Another difference is that their simulations are 3D, while ours are 2D (axisymmetric), which could exaggerate the avalanche accretion stream features and their associated outflows. We note, however, that the 3D simulations of \citet[][]{Zhu2018} clearly show a lightly loaded fast wind that is driven magneto-centrifugally from high-latitude accretion streams. Indeed, our simulations can be viewed as a specialization of their simulations to the young star case that includes a nearly ideal MHD inner disk surrounded by a midplane dead zone at larger radii, with a focus on the innermost part of the simulation domain, which is where most of the gravitational binding energy of the avalanche accretion is released and where the fastest outflow is driven. The presence of a disk midplane dead zone at relatively large radii does not fundamentally change the basic features of the magnetically active regions at smaller radii. 

The prevalence of atmospheric avalanche accretion streams and their associated outflows in simulations similar to ours indicate that they are robust features of magnetized accretion, particularly near the rotation axis where the low density and strong (accumulated) poloidal field combine to enable the upper branch of the stream's pinched field line to magneto-centrifugally accelerate the lightly loaded material to a high speed, forming a protostellar jet.

Although the poloidal magnetic field in the inner active disk zone remains relatively weak in the current generation of simulations, including ours, the question remains whether the poloidal field can become strong enough to disrupt the inner disk through magnetic flux accumulation and drive a fast outflow through accretion across the strong field \citep[similar to magnetically arrested disks (MAD) in black hole accretion, e.g.,][]{Igumenshchev2008}. It is a possibility worth exploring but will be challenging to simulate because of strong magnetization and mass depletion. 

\subsection{Implications and limitations}
\label{subsec:2DLimitation}

The disk, atmosphere, and outflow structure produced in our simulations should affect the transport and thermal processing of dust particles in the inner disk, with implications for the formation of chondrules and calcium- and aluminum-rich inclusions (CAIs). The rather turbulent meridional flows in the active disk zone may lift the dust grains to the magnetically raised atmosphere, where they can be funneled by gas flow into the atmospheric avalanche accretion streams. The raised streams allow the dust grains to be directly exposed to stellar irradiation, facilitating their heating/melting. Perhaps more intriguingly, the frequent magnetic reconnection events in the avalanche streams provide another powerful dust heating/melting mechanism often invoked in chondrule/CAI formation \citep[e.g.,][]{Shu2001}. Just as importantly, some of the molten or partially molten dust droplets can potentially be quickly advected outward by the fast outflow connected to (and launched from) the avalanche streams, where they can cool rapidly, decouple from the outflow, and fall back to the disk at large distances from the central protostar. 

The above picture is broadly similar to the X-wind picture of the chondrule/CAI formation of \citet[][]{Shu2001} but has several notable differences. Firstly, the fast avalanche streams may quickly deliver dust grains from the relatively cool outer regions of the active disk zone outside the dust sublimation radius to the inner disk edge \citep[the ``X-region" in ][]{Shu2001}, which may otherwise be dust-free \citep[][]{Desch2010}. Secondly, the molten dust particles are transported outward by an outflow driven by magnetized disk atmospheric avalanche accretion streams rather than the X-wind driven by the stellar magnetospheric magnetic field near the inner disk edge opened by disk interaction. Furthermore, magnetic reconnections potentially capable of dust-melting occur in the disk atmospheric avalanche streams beyond the inner disk edge rather than in the reconnection ring near the disk midplane interior to the disk inner edge. Whether this variant of \citet[][]{Shu2001}'s picture works quantitatively remains to be determined.

A limitation of the current simulations is the assumption of axisymmetry. This assumption was adopted as a first step to explore efficiently, with reasonable computational costs, the dynamics of the inner disk and jet launching, which requires high spatial resolution and non-ideal MHD effects (i.e., Ohmic dissipation). In particular, high resolution in the $\theta$-direction is required to capture the sharp pinching of the poloidal field lines, which is a key feature of the jet-launching avalanche accretion streams. The assumption of axisymmetry also enabled us to clearly identify reconnecting magnetic field lines using contours of enclosed poloidal magnetic flux. However, some features we observe in the 2D may be modified by 3D effects. 

One feature of the 2D model is the azimuthal magnetic field strength in the active disk zone can be significantly stronger than that in 3D simulations \citep[e.g.][]{Takasao2019, Jacquemin-Ide2021, Zhu2024, Mishra2020, Parfrey2023, Porth2021}. The difference can be attributed to the additional freedom in the azimuthal direction in 3D, allowing easier magnetic field removal through processes unavailable in 2D. For example, magnetic reconnection in the azimuthal direction is only possible in 3D, which can lower the strength of $B_\phi$. Furthermore, magnetic bubbles can form in 3D simulations at some azimuthal locations \citep[e.g.][]{Porth2021}, carrying away poloidal magnetic flux. Because the twist of poloidal magnetic fields generates $B_\phi$, removing poloidal magnetic flux helps reduce the strength of the toroidal magnetic field. Nevertheless, the toroidal magnetic field cannot grow indefinitely in the disk. It is limited by the advection of the toroidal field out of its generation region, which tends to limit the energy density of the toroidal field well below the kinetic energy of the disk rotation, as we demonstrate in the Appendix.

\section{Conclusion}
\label{sec:conclusion}

We carried out a 2D (axisymmetric) non-ideal MHD simulation of the disk around an accreting young stellar object, including a thermally ionized inner active zone surrounded by a magnetically dead zone initially threaded by a relatively weak, large-scale, open poloidal magnetic field. Our main conclusions are as follows.

\begin{itemize}

    \item We find that a fast ($>100$~km/s) collimated bipolar jet is continuously driven along the north and south poles of the circumstellar disk initially magnetized by a large-scale open poloidal field. 
    
    \item The disk-jet system has several distinct components, which, besides the fast-moving jet, include a low-density polar region filled with a strong poloidal magnetic field from magnetic flux accumulation near the axis, a dense equatorial disk that is magnetically active in the inner part surrounded by a magnetically dead zone near the equator at larger radii, a quasi-static atmosphere that is raised above the disk primarily by the toroidal magnetic field and bound to the disk by the stellar gravity, and narrow avalanche accretion streams in the raised atmosphere produced by efficient local magnetic braking. 

    \item The fast jet is primarily driven magneto-centrifugally by the release of the gravitational binding energy of the avalanche accretion streams near the boundary of the evacuated poloidal field-dominated polar region and the denser raised disk atmosphere. In particular, the fast outflow is driven along the upper (open) branch of the highly pinched poloidal field lines threading the accretion streams where the density is relatively low so that the lightly loaded material can be accelerated magneto-centrifugally along the open field line to a high speed. 

    \item We find explicit evidence for repeated reconnection of the highly pinched poloidal magnetic fields threading the avalanche accretion streams, which enables mass to accrete to the center without dragging along the poloidal magnetic flux with it, thus alleviating the problem of excessive magnetic flux accumulation that may disrupt the inner disk and choke the jet launching.  It also provides a potential heating source for producing chondrules and CAIs which, after formation, may then be transported to larger distances by the fast outflow connected to (and driven by) the reconnecting avalanche accretion streams. 
\end{itemize}

\begin{acknowledgments}

We thank Brandt Gaches, Jonathan Tan, and Etienne Martel for helpful discussions at the early stage of this project and the referee for {prompt}  and constructive comments. This work is supported by NASA 80NSSC20K0533 and NSF AST-2307199. Computing resources were provided by the NASA High-End Computing (HEC) Program through the NASA Advanced Supercomputing (NAS) Division at Ames Research Center and the RIVANNA supercomputer at the University of Virginia. YT acknowledges support from an interdisciplinary fellowship from the University of Virginia. 

\end{acknowledgments}

%






\appendix

\section{Limits on the toroidal magnetic field strength in the active disk zone}
\label{sec:activeBlimit}

For disk simulations like ours that start with a relatively weak poloidal magnetic field, the toroidal field is generated by differential twisting of the poloidal magnetic field. We can estimate the strength of the steady-state toroidal field strength from the induction equation under several simplifying assumptions. For this simple estimate, we will assume that the toroial field is generated by the vertical differential twist of a vertical magnetic field $B_z$ and removed by radial advection of a magnetically braked accretion flow. In this case, the induction equation for $B_\phi$ simplifies to 
\begin{equation}
        B_z\frac{\partial v_\phi}{\partial z} \approx  B_\phi\frac{\partial v_R}{\partial R}
    \label{equ:phi_induction}
\end{equation}
where $R$ is the cylindrical radius. We assume that $v_\phi$ is close to the Keplerian value: 
\begin{equation}
    v_\phi(R, z) \approx v_K(R, z) = \frac{(G M)^{1/2} R}{(R^2+z^2)^{3/4}}
    \label{equ:Beq_vphi}
\end{equation}
so that 
\begin{equation}
    \frac{\partial v_\phi(R, z)}{\partial z} = - \frac{3 z\ v_K}{2 (R^2+z^2)}. 
    \label{equ:zderiv}
\end{equation}
We further assume that the radial flow $v_R$ is caused by magnetic braking, which yields: 
\begin{equation}
    v_R = \frac{R}{2\pi\rho v_K}\frac{\partial B_\phi}{\partial z}B_z.
    \label{equ:Beq_vr}
\end{equation}

If the radial variation of $v_R$ is on the scale of radius $R$ and the vertical variation of $B_\phi$ is on the scale of $z_b$, we can combine equations (\ref{equ:phi_induction})-(\ref{equ:Beq_vr}) to eliminate $B_z$ (it cancels out because the toroidal field generation and removal rates both depend linearly on it) and estimate the ratio of the magnetic energy density of the toroidal field and the kinetic energy density from Keplerian rotation
\begin{equation}
    \frac{B_\phi^2/(8\pi)}{\rho \ v_K^2/2} \approx \frac{3z z_b} {4R^2},
    \label{equ:EnergyRatio}
\end{equation}
which is expected to be substantially less than unity inside the disk where both $z$ and $z_b$ are expected to be less than $R$, consistent with what we find in the simulation. 


\bibliography{sample631}{}
\bibliographystyle{aasjournal}



\end{document}